\documentclass[letterpaper,english,reprint, aps, showkeys, showpacs]{revtex4-1}

\usepackage{graphicx}
\usepackage{amsmath}
\usepackage{bbm}

\begin{document}
\title{Elko spinors revised}
\author{R. Romero}
\email{rromero@cua.uam.mx}

\affiliation{Departamento de Ciencias Naturales, División de Ciencias Naturales
e Ingeniería, Universidad Autónoma Metropolitana unidad Cuajimalpa,
Avenida Vasco de Quiroga 4871, Col. Santa Fe Cuajimalpa. Alcaldía
Cuajimalpa de Morelos, C.P. 05348, Ciudad de México.}
\begin{abstract}
It is shown that c-number elko spinors obey the massless Dirac equation
and are unitarily equivalent to Weyl bispinors. Therefore, they do
not constitute a new spinor type with mass dimension one. 
\end{abstract}

\maketitle
Elko spinors are a complete set of $c$-number bispinors that are
eigenstates of the charge conjugation operator, a property from which
they take their name (an acronym from the german \emph{Eigenspinoren
des Ladungs Konjugations Operators}). They were proposed in 2005\cite{PhysRevD.72.067701,1475-7516-2005-07-012}
as the expansion coefficients of a quantum field operator, the elko
field, presented as a new type of fermion field with the exotic properties
of not obeying the massive Dirac equation and having canonical mass
dimension one, despite being fermionic. The latter feature led to
their proposal as a dark matter candidate, since the elko field is
assumed to couple only to the Higgs field to ensure renormalizability.
To this day, elko spinors continue to appear in the scientific literature
in various applications, see e. g. references \citep{Lee2016164,Ahluwalia2017,Villalobos2019,doi:10.1142/S0217732319502110,deBrito2020,Lee2020,AHLUWALIA20221}
and references therein. However, elko spinors are just another type
of massless bispinors satisfying the masslesss Dirac equation, and
as such they can not constitute a new spinor type with mass dimension
different from the known 3/2 mass dimension of fermions in the Standard
Model Lagrangian\cite{pokorski2000gauge}. In fact, they can be unitarily
transformed to massless Weyl spinors, as is shown in this letter.

Let us first state the properties of masless four-component Weyl spinors.
A complete treatment is given in \citep{Romero_2020}, but here we
reproduce the main properties for completeness. Plane wave solutions
to the massless Dirac equation are given by $\Psi=u(\mathbf{p})\exp\left\{ i\left(\pm Et-\mathbf{x}\cdot\mathbf{p}\right)\right\} $,
with the bispinors

\begin{equation}
\begin{array}{cc}
u^{(1)}(\mathbf{p})=\begin{pmatrix}0\\
\chi_{+}\left(\mathbf{p}\right)
\end{pmatrix}, & u^{(2)}(\mathbf{p})=\begin{pmatrix}\chi_{-}\left(\mathbf{p}\right)\\
0
\end{pmatrix},\\
\\
u^{(3)}(\mathbf{p})=\begin{pmatrix}0\\
\chi_{-}\left(\mathbf{p}\right)
\end{pmatrix}, & u^{(4)}(\mathbf{p})=\begin{pmatrix}\chi_{+}\left(\mathbf{p}\right)\\
0
\end{pmatrix},
\end{array}\label{eq:2.10-1}
\end{equation}

\noindent and the two-component spinors $\chi_{\pm}\left(\mathbf{p}\right)$
given by

\begin{align}
\begin{split}\chi_{+}\left(\mathbf{p}\right)= & \begin{pmatrix}\cos\left(\frac{\theta}{2}\right)\\
e^{i\varphi}\sin\left(\frac{\theta}{2}\right)
\end{pmatrix},\\
\chi_{-}\left(\mathbf{p}\right)= & \begin{pmatrix}-e^{-i\varphi}\sin\left(\frac{\theta}{2}\right)\\
\cos\left(\frac{\theta}{2}\right)
\end{pmatrix}.
\end{split}
\label{eq:2.11-1}
\end{align}

\noindent For definiteness, we use the gamma matrices Weyl representation,
with the following definitions

\begin{equation}
\begin{array}{c}
\gamma^{0}=\left(\begin{array}{rr}
0 & 1\\
1 & 0
\end{array}\right),\,\boldsymbol{\gamma}=\left(\begin{array}{rr}
0 & \boldsymbol{\sigma}\\
-\boldsymbol{\sigma} & 0
\end{array}\right),\end{array}\label{eq:2.2-1}
\end{equation}

\begin{equation}
\begin{array}{c}
\gamma^{5}\equiv i\gamma^{0}\gamma^{1}\gamma^{2}\gamma^{3}=\left(\begin{array}{rr}
0 & 1\\
1 & 0
\end{array}\right),\,\boldsymbol{\Sigma}\equiv\gamma^{5}\gamma^{0}\boldsymbol{\gamma}=\left(\begin{array}{rr}
\boldsymbol{\sigma} & 0\\
0 & \boldsymbol{\sigma}
\end{array}\right),\end{array}\label{eq:2.4-1}
\end{equation}

\noindent where $\boldsymbol{\sigma}=\left(\sigma^{1},\sigma^{2},\sigma^{3}\right)$
are the standard Pauli matrices. Then the massless Dirac equation
$i\gamma^{\mu}\partial_{\mu}\Psi=0$ simplifies to 
\begin{equation}
\boldsymbol{\Sigma}\cdot\mathbf{\hat{p}}\,u(\mathbf{p})=\pm\gamma^{5}\,u(\mathbf{p}).\label{eq:2.5-1}
\end{equation}

In Hamiltonian form Eq.(\ref{eq:2.5-1}) reads

\begin{equation}
\begin{split}\boldsymbol{\alpha}\cdot\hat{\mathbf{p}}\,u^{(s)}(\mathbf{p})= & +u^{(s)}(\mathbf{p}),\\
\boldsymbol{\alpha}\cdot\hat{\mathbf{p}}\,u^{(s+2)}(\mathbf{p})= & -u^{(s+2)}(\mathbf{p}),
\end{split}
s=1,2\label{eq:2.13-1}
\end{equation}

\noindent and $u^{(1)}(\mathbf{p})$ and $u^{(2)}(\mathbf{p})$ are
positive-energy bispinors, with both positive helicity an chirality
for the former and negative for the latter, while $u^{(3)}(\mathbf{p})$
and $u^{(4)}(\mathbf{p})$ are negative-energy ones, with negative
helicity and positive chirality for the former and the reversed values
for the latter. These bispinors are orthonormal $\left[u^{(i)}(\mathbf{p})\right]^{\dagger}u^{(j)}(\mathbf{p})=\delta_{ij}$.
Taking the momentum $\hat{\boldsymbol{p}}=\left(\sin\theta\cos\varphi,\sin\theta\sin\varphi,\cos\theta\right)$
in the $\hat{\mathbf{z}}$ direction, which will be referred to as
the canonical frame, the bispinors simplify to

\begin{equation}
\begin{array}{rr}
u^{(1)}\left(p_{z}\right)=\begin{pmatrix}0\\
0\\
1\\
0
\end{pmatrix}, & u^{(2)}\left(p_{z}\right)=\begin{pmatrix}0\\
1\\
0\\
0
\end{pmatrix},\\
u^{(3)}\left(p_{z}\right)=\begin{pmatrix}0\\
0\\
0\\
1
\end{pmatrix}, & u^{(4)}\left(p_{z}\right)=\begin{pmatrix}1\\
0\\
0\\
0
\end{pmatrix}.
\end{array}\label{eq:2.6}
\end{equation}

\noindent The canonical frame bispinors and the general momentum ones
are related by the rotation

\begin{equation}
\Lambda_{1}\left(\theta,\varphi\right)=\exp\left\{ -\dfrac{\theta}{2}\left(\gamma^{1}\cos\varphi+\gamma^{2}\sin\varphi\right)\gamma^{3}\right\} ,\label{eq:2.8}
\end{equation}

\noindent since

\begin{equation}
\begin{array}{ccc}
\Lambda_{1}\left(\theta,\varphi\right)u^{(i)}\left(p_{z}\right)=u^{(i)}(\mathbf{p}), &  & i=1,\ldots,4.\end{array}\label{eq:2.9}
\end{equation}

Let us now define the elko bispinors following reference \onlinecite{AHLUWALIA20221}.
These are

\begin{equation}
\begin{array}{cc}
\lambda^{(1)}(\mathbf{p})=\begin{pmatrix}\sigma^{2}\eta_{+}^{*}(\mathbf{p})\\
\eta_{+}(\mathbf{p})
\end{pmatrix}, & \lambda^{(2)}(\mathbf{p})=\begin{pmatrix}\sigma^{2}\eta_{+}^{*}(\mathbf{p})\\
-\eta_{+}(\mathbf{p})
\end{pmatrix},\\
\\
\lambda^{(3)}(\mathbf{p})=\begin{pmatrix}-\sigma^{2}\eta_{-}^{*}(\mathbf{p})\\
\eta_{-}(\mathbf{p})
\end{pmatrix}, & \lambda^{(4)}(\mathbf{p})=\begin{pmatrix}\sigma^{2}\eta_{-}^{*}(\mathbf{p})\\
\eta_{-}(\mathbf{p})
\end{pmatrix},
\end{array}\label{eq:2.25}
\end{equation}

\noindent where we have changed the notation and ignore the ad-hoc
normalization used in the aforementioned reference, for simplicity.
The two-component spinors $\eta_{\pm}(\mathbf{p})$ are given by

\begin{equation}
\begin{split}\eta_{+}\left(\mathbf{p}\right)= & \begin{pmatrix}e^{-i\varphi/2}\cos\left(\frac{\theta}{2}\right)\\
e^{i\varphi/2}\sin\left(\frac{\theta}{2}\right)
\end{pmatrix},\\
\eta_{-}\left(\mathbf{p}\right)= & \begin{pmatrix}-e^{-i\varphi/2}\sin\left(\frac{\theta}{2}\right)\\
e^{i\varphi/2}\cos\left(\frac{\theta}{2}\right)
\end{pmatrix},
\end{split}
\label{eq:2.26}
\end{equation}

\noindent which differ from the spinors in Eq. (\ref{eq:2.11-1})
by a phase

\begin{equation}
\chi_{\pm}=e^{\pm i\varphi/2}\eta_{\pm}.\label{eq:2.27}
\end{equation}

\noindent The elko bispinors are eigenstates of the charge conjugation
operator $\mathcal{C}\equiv\gamma^{2}\mathcal{K}$, which is their
defining property, with $\mathcal{K}$ representing complex conjugation
to the right

\begin{equation}
\begin{split}\mathcal{C}\lambda^{(1,4)}(\mathbf{p})= & +\lambda^{(1,4)}(\mathbf{p}),\\
\mathcal{C}\lambda^{(2,3)}(\mathbf{p})= & -\lambda^{(2,3)}(\mathbf{p}).
\end{split}
\label{eq:2.28}
\end{equation}

\noindent Now, a straightforward calculation shows that the elko bispinors
are solutions to the massless Dirac equation

\begin{equation}
\begin{split}\boldsymbol{\alpha}\cdot\hat{\mathbf{p}}\,\lambda^{(s)}(\mathbf{p})= & +\lambda^{(s)}(\mathbf{p}),\\
\boldsymbol{\alpha}\cdot\hat{\mathbf{p}}\,\lambda^{(s+2)}(\mathbf{p})= & -\lambda^{(s+2)}(\mathbf{p}).
\end{split}
s=1,2,\label{eq:2.36-1}
\end{equation}

\noindent Hence, the correct field operator, expanded in terms of
these spinors, would necessarily be that of a massless Dirac field,
satisfying the massless Dirac equation, as Weinberg has shown in a
seminal paper\cite{PhysRev.134.B882}. Furthermore, the associated
propagator, either for the spinors in a Relativistic Quantum Mechanics
framework or for the field operator in Quantum field Theory, would
have to be the masless Dirac propagator. Therefore, a massive spin
$1/2$ field operator in terms of elko spinors, as defined in reference
\onlinecite{AHLUWALIA20221}, with mass dimension one and that does
not obey the massive Dirac equation is just unphysical.

Having proved that Elko spinors obey the massless Dirac equation,
we now have that both Elko and Weyl bispinors satisfy an eigenvalue
equation with the same Hamiltonian and $\pm$ eigenvalues, as shown
in Eqs. (\ref{eq:2.13-1}) and (\ref{eq:2.36-1}). Let us schematically
write them as $\mathcal{H}u(\mathbf{p})=\pm u(\mathbf{p})$ and $\mathcal{H}\lambda(\mathbf{p})=\pm\lambda(\mathbf{p})$,
with $\mathcal{H}=\boldsymbol{\alpha}\cdot\hat{\mathbf{p}}$ the massless
Dirac Hamiltonian. Then, there must be a unitary transformation $\Omega\left(\theta,\varphi\right)$
with the properties $\Omega\lambda(\mathbf{p})=u(\mathbf{p})$ and
$\Omega\mathcal{H}-\mathcal{H}\Omega=0$, such that $\Omega\mathcal{H}\Omega^{-1}\Omega\lambda(\mathbf{p})=\pm\Omega\lambda(\mathbf{p})$
implies $\mathcal{H}u(\mathbf{p})=\pm u(\mathbf{p})$

To this end let us consider the rotation

\begin{align}
\begin{split}\Lambda_{2}\left(\theta,\varphi\right)= & \sin\left(\frac{\varphi}{2}\right)\left[\cos\left(\frac{\theta}{2}\right)\gamma^{1}+\sin\left(\frac{\theta}{2}\right)\gamma^{3}\right]\gamma^{2}\\
+ & \cos\left(\frac{\varphi}{2}\right)\left[\cos\left(\frac{\theta}{2}\right)\mathbbm{1}-\sin\left(\frac{\theta}{2}\right)\gamma^{1}\gamma^{3}\right],
\end{split}
\label{eq:2.29}
\end{align}

\noindent which transforms the elko bispinors from the canonical frame
to the general momentum bispinors in Eq. (\ref{eq:2.25}), that is

\begin{equation}
\begin{array}{ccc}
\Lambda_{2}\left(\theta,\varphi\right)\lambda{}^{(i)}\left(p_{z}\right)=\lambda^{(i)}(\mathbf{p}), &  & i=1,\ldots,4,\end{array}\label{eq:2.30}
\end{equation}

\noindent where the $\lambda^{(i)}\left(p_{z}\right)$ correspond
to the $\lambda^{(i)}(\mathbf{p})$ with $\theta=\varphi=0$. The
canonical frame elko bispinors can be rotated to the corresponding
Weyl bispinors, by means of the rotation

\begin{align}
\begin{split}U= & \frac{i}{2\sqrt{2}}\left(-i\left(\mathbbm{1}+\gamma^{5}\right)+\gamma^{1}\left(\mathbbm{1}+i\gamma^{2}\right)\right.\\
+ & \left.\gamma^{2}-\left(\mathbbm{1}-i\left(\gamma^{1}+\gamma^{2}\right)\right)\gamma^{0}\gamma^{3}\right),
\end{split}
\label{eq:2.31}
\end{align}

\noindent yielding

\begin{equation}
\begin{array}{cc}
U\lambda^{(1)}\left(p_{z}\right)=u^{(1)}\left(p_{z}\right), & U\lambda^{(2)}\left(p_{z}\right)=u^{(2)}\left(p_{z}\right),\\
U\lambda^{(3)}\left(p_{z}\right)=u^{(3)}\left(p_{z}\right), & U\lambda^{(4)}\left(p_{z}\right)=u^{(4)}\left(p_{z}\right).
\end{array}\label{eq:2.32}
\end{equation}

\noindent Defining the rotation

\begin{equation}
\Omega\left(\theta,\varphi\right)\equiv\Lambda_{1}\left(\theta,\varphi\right)U\Lambda_{2}^{\dagger}\left(\theta,\varphi\right),\label{eq:2.33}
\end{equation}

\noindent we obtain from Eqs. (\ref{eq:2.9}), (\ref{eq:2.30}) and
(\ref{eq:2.32}), a relation between general momentum elko and massless
Weyl bispinors

\begin{equation}
\begin{array}{cc}
\Omega\left(\theta,\varphi\right)\lambda^{(1)}\left(\boldsymbol{p}\right)=u^{(1)}\left(\boldsymbol{p}\right), & \Omega\left(\theta,\varphi\right)\lambda^{(2)}\left(\boldsymbol{p}\right)=u^{(2)}\left(\boldsymbol{p}\right),\\
\Omega\left(\theta,\varphi\right)\lambda^{(3)}\left(\boldsymbol{p}\right)=u^{(4)}\left(\boldsymbol{p}\right), & \Omega\left(\theta,\varphi\right)\lambda^{(4)}\left(\boldsymbol{p}\right)=u^{(3)}\left(\boldsymbol{p}\right).
\end{array}\label{eq:2.34}
\end{equation}

\noindent It can also be shown that, as required,

\begin{equation}
\left[\Omega\left(\theta,\varphi\right),\boldsymbol{\alpha}\cdot\hat{\mathbf{p}}\right]=0.\label{eq:2.35}
\end{equation}

\noindent Therefore, elko spinors can be obtained from Weyl bispinors
by a rotation and vice versa, which is another way to show that the
former do not constitute a new type of spinors, but are in fact equivalent
to massless Weyl bispinors, obeying the massless Dirac equation.

\bibliographystyle{apsrev4-2}

\end{document}